\input aa.cmm
\input epsf
\MAINTITLE={On the possibility of curvature radiation from radio pulsars}
\AUTHOR={H. Lesch@1, A. Jessner@2, M. Kramer@2 and T. Kunzl@1 }
\OFFPRINTS={H. Lesch}
\INSTITUTE={@1 Institut f\"ur Astronomie und Astrophysik, Universit\"at
M\"unchen, Scheinerstr. 1, 81679 M\"unchen, FRG;

@2 Max Planck-Institut f\"ur Radioastronomie, 53121 Bonn, Auf dem
H\"ugel 69, FRG
}
\DATE={}
\ABSTRACT={
We consider the widespread hypothesis that coherent curvature
radiation is responsible for the radio emission of pulsars.
The comparison of energy conservation and the published data and
luminosities explicitely proves that coherent curvature radiation {\bf
cannot} be the source for the radio emission of pulsars for
frequencies below a few GHz. At higher frequencies coherent curvature
radiation can be ruled out because 
neither the observationally deduced emission heights nor the observed
radius to frequency mapping can be reproduced by this mechanism.
Our argumentation is in accordance with the more general critics
(e.g. Melrose 1992) that no adequate bunching mechanism has been
identified for coherent curvature radiation. 
We present 5 examples (0329+29, 0355+54, 0540+23, 1133+16, 1916+10)
of pulsars whose high frequency (larger than 1.4. GHz, up to 32 GHz)
luminosities are well known, and as a low frequency example
the faintest radio pulsar 0655+64 of the Taylor et al (1993) sample.}
\KEYWORDS={Plasmas - Radiation mechanism - Pulsars}
\THESAURUS={02.16.1; 02.18.5; 08.16.6}
\maketitle

\titlea{Introduction}

Pulsar radio emission has a very high brightness temperature which implies
that the emission mechanism must be coherent, that is the emission cannot
be explained in terms of individual particles radiating independently
(incoherently) of each other. There are several models for the origin
of such a coherent mechanism (e.g. Asseo 1993 and references
therein). Here we concentrate on one of the earliest and still most
prominent models, namely coherent curvature radiation (Gunn and
Ostriker 1971; Sturrock 1971; Ruderman and Sutherland 1975; Ginzburg
and Zheleznyakov 1975; Buschauer and Benford 1976; Kirk 1980;
Buschauer and Benford 1983). In the frame of a pulsar as a rotating
dipole it is natural to consider the radiation of relativistic charged
particles when they move along the curved field lines, thereby
emitting curvature radiation. The necessary bunching mechanism was
supposed to be provided by a two-stream instability, which excites
plasma waves. They are supposed to bunch particles via their
electrostatic fields (e.g. Ruderman and Sutherland 1975). This ansatz
has been criticized for several reasons (Melrose 1992 and references
therein) A fundamental difficulty is that the theory for bunching
instabilities does not allow for any velocity dispersion of the
emitting particles. No adequate bunching mechanism has been
identified. An extreme from of bunching is required; specifically the
relativistic particles of single sign need to form well separated
bunches with a pancake shape with normal almost exactly along the
field lines. It was concluded, that in view of these difficulties
coherent curvature radiation should not be regarded as the favored
mechanism for pulsar radio emission (Asseo et al 1980, 1983; Melrose
1992)). 

Despite this irrefutable theoretical critics coherent curvature
radiation is still one of the favorites of many phenomenological
models (Rankin 1983a,b, 1986, 1988, 1989, 1992; Radhakrishan and
Rankin 1990; Gil 1983, 1984, 1992; Gil and Snakowski 1990a,b). 
As a step towards a clarification we present here a simple argument
against coherent curvature radiation as a source for radio emission
from pulsar by means of energy conservation. We investigate whether
coherent curvature radiation is able to produce the observed
luminosities within the constraints of emission heights derived from
the observations. Thus, we do not consider the criticism about the
origin of bunches. We ask: is it possible to release enough power
in order to explain the observed luminosities via coherent curvature
radiation? 

\titlea{Coherent Curvature Radiation}
Curvature radiation can be described in terms of emission
by a relativistic particle moving around the arc of a circle
chosen such that the actual acceleration corresponds to centripetal
acceleration.
A relativistic electron with Lorentz factor $\gamma$, constrained to
follow a path with radius of curvature $R_{\rm c}$, radiates similarly
to an electron in a circular orbit with frequency  $c/2\pi R_{\rm
c}$. The critical frequency where most of the radiation is emitted is
approximately given by (e.g. Zheleznyakov 1996, p. 231)

$$\nu_{\rm c}\simeq {3c\over{4\pi R_{\rm c}}}\gamma^3.\eqno(1a)$$
Or if we observe a particular frequency we find the corresponding
Lorentz factor via 

$$\gamma\simeq \left[{\nu_{\rm c} 4\pi R_{\rm
c}\over{3c}}\right]^{1/3}.\eqno(1b)$$ 

The curvature radius of the field lines in a magnetic dipole field can
be expressed as (Smirnow 1973, p. 211)

$$ R_c(\theta,\theta_s)={
{r_{ns}\over 3}\cdot
{\sin\theta \over \sin^2\theta_s}\cdot {{(1+3\cos^2\theta)^{3/2}}
\over {1+\cos^2\theta} } 
}
\eqno(2) $$

where $\theta$ is the colateral angle of a point on the field line
starting at $\theta_s$ at the pulsar surface. A reasonable
approximation up to $50\cdot r_{ns}$ is given by  
$$R_{c}(r)\simeq R_0\cdot \sqrt{{r\over{r_{ns}}}},\eqno(3)$$

where $R_0 = R_c(r_{ns}) =R_c(\theta_s,\theta_s)$ is the curvature
radius at the pulsar surface. 

The total power radiated by a particle of energy $E=\gamma m_e c^2$
along a curved field line 
is ($q_e$ denotes the charge)

$$P_{\rm c}={2\over 3}\gamma^4 {q_e^2 c\over{4\pi \epsilon_0
R_{c}^2}}.\eqno(4)$$ 

The classical Goldreich-Julian charge density in the pulsar
magnetosphere (Goldreich and Julian 1969)

$$ n_{GJ}= {{2 \epsilon_0\Omega B_0}\over q_e}\cdot {r_{ns}^3\over r^3}
\eqno(5)  $$

provides us with a typical particle density of $n_{GJ}(r_{ns})=
3.6\cdot 10^{11} cm^{-3}$ if P=0.5 s and $B=10^{12}$ G.
It enables us to estimate
the typical size of the emission region for {\it incoherent curvature
emission}. For the lowest luminosity $L=10^{25} erg/s$ (0655+64
(Taylor et al. 1993) we find $({L\over{P_c\cdot n_{GJ}}})^{1/3} = 4000
km$. Using equations 1b, 4 and 5 together with the simplifying
assumption that the emission region extends across the line of sight
as well as along it, we find that the predicted profile widths for
incoherent curvature emission turn out to be within 

$$w_{inc}(\nu) =2\cdot \tan^{-1}\Bigl(\bigl({81 c R_c^2 \over 4 \pi
\nu^4}\bigr)^{1/9} 
\bigl({{3 L} \over {4 q_e \Omega B_0}}\bigr)^{1/3}\Bigr)  \eqno(6)$$

For frequencies below $10^{16}\ Hz$ we predict profile widths of
roughly 180 degrees and only when we approach  $10^{19}\ Hz$ ($h\nu =
4 keV$) can we expect widths of around 25 degrees. Typical pulsar
profile widths at radio frequencies are however of the order of 5-10
degrees (e.g. Izvekova et. al. 1994, Seiradakis et. al. 1995). 
Because of the narrow profiles of the pulsed emission we must rule out
any consideration of {\it incoherent curvature emission} as the source
of the observed pulsar radio luminosities.

As incoherent emission is obviously insufficient it might be
worthwhile to turn or attention to {\it coherent} curvature emission
which has been suggested as an emission mechanism (see references
above). Coherence exists in volumes of the order of $V_c =
c^3\gamma^2/\pi\nu^3$ (Melrose 1992).
The power output of such a coherence volume is then given by:

$$P_{N}={2\over 3}\gamma^4 {(N_c\cdot q_e)^2 c\over{4\pi \epsilon_0
R_{c}^2}} = P_c\cdot N_c^2 \eqno(7)$$ 

with $N_c=V_c\cdot n_{gj}$ being the number of particles within a
coherence volume. 
To account for the observed luminosity L one requires $N_v ={L\over
P_c\cdot N_c^2} = 7\cdot 10^{11}$ of these volumes, which will give us
an estimate of the total number of particles involved

$$N = N_c\cdot N_v = {L \over P_c\cdot N_c }  \eqno(8)$$

and the minimal size of the emission region

$$\Lambda_c = \left(N_v\cdot V_c\right)^{1/3}  \eqno(9) $$

Energy conservation alone provides us with an important constraint
that will enable us to restrict the free parameter $\gamma$ and to
check the consistency of the model.

The energy loss of a charge within the emission region is now

$$ P_\Lambda={\Lambda_c\over c}P_c N_c \eqno(10)$$
which must never exceed the energy of the particle $\gamma m_e c^2$
itself.

The bunching of particles requires an external electric force which
confines the particles in a coherence volume.  Although electrostatic
instabilities have been criticised for several reasons, the principal
argument, that electrostatic plasma wave can be responsible for the
bunching of particles has not been questioned, only processes 
that excite the waves are under discussion. The Doppler shifted plasma
waves oscillate with the local plasma frequency
$\omega_{\rm pe}=\gamma\sqrt{{n q_e^2\over{\epsilon_0 \gamma m_e}}}$
of the particles and therefore their electrostatic
fields present the ideal bunching force. When the particles
are in resonance with the wave  they experience the wave electric
field most intensively and they are confined (bunched) in the
coherence volume 
$\propto \lambda^3\propto (c/\omega_{\rm pe})^3$.
The same resonance condition
holds for the propagation of electromagnetic waves  with frequency
$\nu$ (where $\nu\ge \omega_{\rm pe}/2\pi$), which defines a certain
radius in the pulsar magnetosphere, at which the coherent radiation
can escape from the plasma.
Particle bunching and wave propagation is determined by the same
condition $\nu\ge \omega_{\rm pe}/2\pi$ (e.g. Melrose 1992). Since the
bunching leads to a very fast and efficient emission the place at
which the condition is fulfilled is also the place of emission,
i.e. the emission height. For larger radii only lower frequencies can
be emitted (see eqs. 1a and 3). An increase in density corresponds 
to an increase in frequency at a given radius (eq. 5).
With the determination of the minimum emission height from
$\nu\ge \omega_{\rm pe}/2\pi$ using eq.(1b) to eliminate $\gamma$

$$ \nu = {3^{2/5}\over {4 \pi c^{1/5} } } \Bigl({{8 q_e \Omega B_0
r_{ns}^3}\over{3 m_e}} \Bigr)^{3/5}r^{-9/5}R_c^{1/5} \eqno(11) 
$$

we can express the ratio of emitted energy over the total energy as

$$ \eta_1 = {P_\Lambda \over \gamma m_e c^2 }
\eqno(12)$$

which cannot exceed unity. From the condition $\eta_1\leq 1$ we obtain
a condition for the luminosity by inserting eqs.(1-9) into eq. (10),
leaving 

$$ L_{\rm crit}={\gamma^3 m_e^3 c^9\over{V_{\rm c} P_{\rm c}^2 N_{\rm
c}}}= {243 \over 32} {{\pi \epsilon_0 m_e^3 c^2}\over{q_e^3 \Omega
B_0}}\nu^3 {r^3 \over r_{ns}^3} R_c \eqno(13)$$

\titlea{Application to six pulsars}
Clearly, the observed luminosities must not exceed $L_{\rm crit}$
if coherent curvature radiation is to be a dominant mechanism.
We can now compare the observed luminosities at different frequencies
with the corresponding critical luminosity provided by coherent curvature
radiation. For frequencies up to 32 GHz we use the sample of Kramer et
al. (1996): 0329+29;0355+54; 0540+23; 1133+16; 1916+10. As a limiting
case we use also the weakest pulsar in the cataloque of Taylor et
al. (1993): 0655+64 which has a maximal luminosity at 408 MHz of
$3\cdot 10^{25}\, erg s^{-1}$. 

\epsfxsize=\hsize
\epsfbox{curv1.prn}
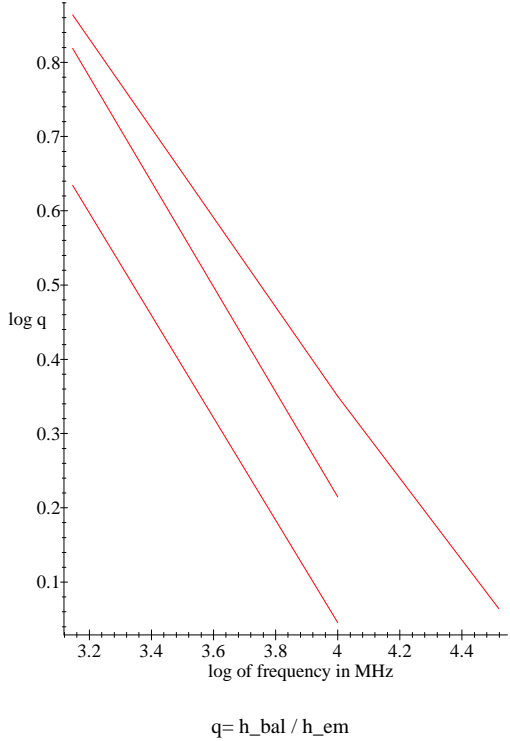
\figure{1} {$h_{bal}/h_{em}$ versus frequency for
three pulsars (from left to right: 1929+10; 0540+32; 0329+54). q>1
means that curvature radiation can \bf{not} provide enough energy at
the corresponding frequency}

Energetically, curvature radiation is not ruled out wherever $L_{crit}
> L_{observed}$. We will call the minimal surface distance for which
we would have $L_{crit}/L_{observed} > 1$ the energy balance distance
$h_{bal}$ to contrast it from the minimal emission distance
$h_{em}$,-- found by inverting  eq. (11), where the curvature
radiation frequency matches the plasma frequency. We present the
ratio $h_{bal}/h_{em}$ versus frequency for five pulsars in
Fig. 1. Apart from the highest frequencies we find that it is
impossible to fulfill the condition $h_{em}/h_{bal}<1$ for frequencies
below several GHz.  

\epsfxsize=\hsize
\epsfbox{curv2.prn}
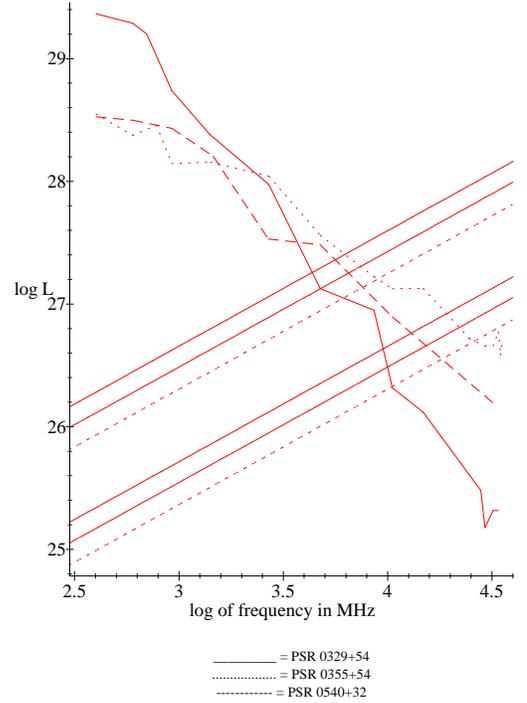
\figure{2a}{Luminosities of the three pulsars mentioned, compared with
the maximum possible luminosity  that can be obtained by coherent
curvature radiation. The upper straight lines are calculated with
$n=10^4 n_{GJ}$, whereas for the lower ones $n=n_{GJ}$ is assumed.} 

\titlea{Conclusions}

We note that nearly everwhere the theoretical emission heights are too
large if compared with the corresponding observational estimates,
(Kramer et al. 1996; Kijak and Gil 1997). This is particularly
striking when one considers the  frequency regime below 1.4 GHz. 
Here the observations often yield luminosities that cannot be achieved anywhere
within the bounds of pulsar magnetosphere. As it is unlikely that all
of the beam energy is going to be emitted as curvature radiation we
have to realise that the actual emission heigths would even be greater
than from these simple estimates. Furthermore,
from eq. (3) and (11) we obtain a radius to frequency mapping exponent
of $\chi=-10/17=-0.59$, whereas the observed exponent is $-0.15\pm
0.1$ (Gil and Kijak 1992; Kramer et al. 1996). 
The radius to frequency method determines the frequency dependent
radius $R(\nu)$, at which 
the emission should originate:
$R(\nu)\simeq R_0 +C\cdot\nu^{\chi}$ (C is a constant with dimension
$m Hz^\chi$ (Phillips 1992).

Although at high frequencies (above 10 GHz) $L_{crit}$ exceeds
$L_{observed}$ by many orders of magnitude, (i.e. there is no
energetic argument to exclude curvature radiation), we rule out
coherent curvature radiation since it neither reproduces the observed
radius to frequency mapping nor the estimated emission heights.

\epsfxsize=\hsize
\epsfbox{curv3.prn}
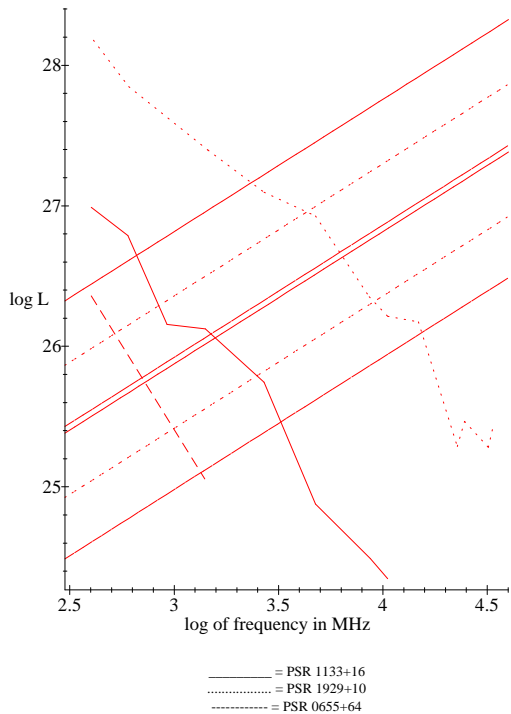
\figure{2b} {The same as Fig. 2a, but for three other pulsars.}

In Fig.2a and 2b we show the frequency dependent $L_{\rm crit}$ given
by eq. (13) versus frequency for the 6 pulsars with different
magnetospheric densities. The solid lines are for $n=n_{\rm GJ}$ and
the dashed lines for $n=10^4 n_{\rm GJ}$. The latter case is used
since several models for $\gamma$-radiation involve massive pair 
production above the polar cap region, i.e. much more secondary
particles are produced than primary particles (Daugherty and Harding
1994, 1996; Usov and Melrose 1996; Zhang et al. 1997; Miyazaki and
Takahara 1997). Obviously even for the high density
case the observed luminosities at frequencies lower than about 6 GHz
cannot be explained by coherent curvature radiation and since it
cannot explain emission heights and radius to frequency mapping, we
exclude this mechanism as origin for pulsar radio emission.

Other mechanisms like for example nonlinear electrostatic
instabilities which drive strong Langmuir turbulence, as proposed by
Asseo (1993) should be considered as promising alternatives to
coherent curvature radiation at low radio frequencies where the bulk
of the radioemission is received. A comparison with this mechanism and
the observations will be shown in a following paper.

\titlea{References}
\ref Asseo, E., 1993, MNRAS 264, 940
\ref Asseo, E., Pellat, R., Rosado, M., 1980, ApJ 239, 661
\ref Asseo, E., Pellat, R., Sol, H., 1983, ApJ 266, 201
\ref Benford, G., Buschauer, R., 1983, A\&A 118, 358
\ref Buschauer, R., Benford, G., 1976, MNRAS 177, 109
\ref Daugherty, J.K., Harding, A.K., 1994, ApJ 429, 325
\ref Daugherty, J.K., Harding, A.K., 1996, ApJ 458, 278
\ref Gil, J.A., 1983, A\& A 123, 7
\ref Gil, J.A., 1984, 1984, A\& A 131, 67
\ref Gil, J.A., 1992,  in {\it The Magnetospheric Structure and
Emission Mechanisms of Radio Pulsars}, eds. 
Hankins, T.H., Rankin, J.M. and Gil, J.A., IAU Colloq. {\bf 128}, p. 395
\ref Gil, J.A., Snakowski, J.K., 1990a, A\& A 234, 237
\ref Gil, J.A., Snakowski, J.K., 1990b, A\& A 234, 269
\ref Gil, J.A., Kijak, J., 1992, A\& A 256, 477
\ref Ginzburg, V.L., Zheleznyakov, V.V., 1975, ARAA 13, 511
\ref Goldreich, P., Julian, W.H., 1969, ApJ 157, 869
\ref Gunn, J.E., Ostriker, J.P., 1971, ApJ 165, 523
\ref Izvekova, V.A., Jessner, A., Kuzmin, A.D., et al., 1994,
A\&A Suppl. 105,235
\ref Kijak, J., Gil, J., 1997, MNRAS (in press)
\ref Kirk, J.G., 1980, A\&A 82, 262
\ref Kramer, M. Xilouris, K.M., Jessner, A., et al., 1996,
A\&A 306, 867
\ref Melrose, D.B., 1992,  in {\it The Magnetospheric Structure and
Emission Mechanisms of Radio Pulsars}, eds. 
Hankins, T.H., Rankin, J.M. and Gil, J.A., IAU Colloq. {\bf 128}, p. 306
\ref Michel, F.C., 1978,  ApJ 220, 1101
\ref Miyazaki, J.F., Takahara, F., 1997, MNRAS 290, 49
\ref Phillips, J.A., 1992, ApJ 385, 282
\ref Radhakrishan, V., Rankin, J.M., 1990, ApJ 352, 258
\ref Rankin, J.M., 1983a, ApJ 274, 333
\ref Rankin, J.M., 1983b, ApJ 274, 359
\ref Rankin, J.M., 1986, ApJ 301, 901
\ref Rankin, J.M., 1988, ApJ 325, 314
\ref Rankin, J.M., 1990, ApJ 352, 247
\ref Rankin, J.M., 1992,  in {\it The Magnetospheric Structure and
Emission Mechanisms of Radio Pulsars}, eds. 
Hankins, T.H., Rankin, J.M. and Gil, J.A., IAU Colloq. {\bf 128},
p. 133 
\ref Ruderman, M.A., Sutherland, P.G., 1975, ApJ 196, 51
\ref J.H. Seiradakis, J.A. Gil, D.A. Graham, et al., 1995, A\&A
Suppl., 111, 205-227 
\ref Smirnow, W.I., 1973 {\it Lehrgang der h\"oheren Mathematik},
Akade\-mie-Verlag (Berlin) 
\ref Taylor, J.H., Manchester, R.N., Lyne, A.G., 1993, ApJS, 88, 529
\ref Usov, V.V., Melrose, D.B., 1996, ApJ 464, 306
\ref Zhang, B., Qiao, G.J., Lin, W.P., Han, J.L., 1997, ApJ 478, 313
\ref Zheleznyakov, V.V.  1996 {\it Radiation Processes in
Astrophysical Plasmas}, Kluwer, Dordrecht, p. 231

\bye